\documentclass[journal]{IEEEtran}
 \pdfoutput=1
 \usepackage{cite}
\usepackage{amsmath,amssymb,amsfonts}
\usepackage{algorithmic}
\usepackage{graphicx}
\usepackage{textcomp}
\usepackage{framed,multirow}
\usepackage{latexsym}
\usepackage{amsmath}
\usepackage{threeparttable}
\usepackage{float}
\usepackage{array}
\usepackage{hhline}
\usepackage{colortbl}
\usepackage{stfloats}
\usepackage{booktabs}
\usepackage{multirow}
\usepackage{enumerate}
\usepackage{diagbox}
\usepackage{url}
\usepackage{hyperref}
\usepackage{makecell,rotating}

\def\etal{\textit{et~al}.}

\DeclareGraphicsExtensions{.pdf,.jpeg,.png,.jpg}

\graphicspath{{fig/}}

\definecolor{mygray}{gray}{.9}
\newcommand{\red}[1]{\textcolor{red}{#1}}
\newcommand{\green}[1]{\textcolor{green}{#1}}
\newcommand{\blue}[1]{\textcolor{blue}{#1}}
\definecolor{newcolor}{rgb}{.8,.349,.1}

\makeatletter

\newcommand{\Rmnum}[1]{\expandafter\@slowromancap\romannumeral #1@}
\makeatother

\title{CKD-TransBTS: Clinical Knowledge-Driven Hybrid Transformer with Modality-Correlated Cross-Attention for Brain Tumor Segmentation}
\author{Jianwei Lin, Jiatai Lin, Cheng Lu, Hao Chen, Huan Lin, Bingchao Zhao, Zhenwei Shi, Bingjiang Qiu, Xipeng Pan, Zeyan Xu, Biao Huang, Changhong Liang, Guoqiang Han, Zaiyi Liu, Chu Han, \IEEEmembership{Member, IEEE}
\thanks{This work was supported by 
the Key R\&D Program of Guangdong Province, China (No. 2021B0101420006), 
the National Key R\&D Program of China (No. 2021YFF1201003), 
the National Science Fund for Distinguished Young Scholars (No.81925023), 
the National Science Foundation for Young Scientists of China (No. 62102103, 62002082 and 82102034),
the National Natural Science Foundation of China (No. 82071892 and 82071871), 
High-level Hospital Construction Project (No. DFJH201805 and DFJH201914), 
China Postdoctoral Science Foundation (No. 2021M690753 and 2022M710843).
(Corresponding author: Zaiyi Liu, Guoqiang Han, and Chu Han.)}
\thanks{Jianwei Lin, Jiatai Lin and Guoqiang Han are with the School of Computer Science and Engineering, South China University of Technology, Guangzhou, 510006, China.}
\thanks{Cheng Lu, Huan Lin, Bingchao Zhao, Zhenwei Shi, Bingjiang Qiu, Xipeng Pan, Zeyan Xu, Biao Huang, Changhong Liang, Zaiyi Liu and Chu Han are with the Department of Radiology, Guangdong Provincial People’s Hospital, Guangdong Academy of Medical Sciences, Guangzhou, 510080, China;
Guangdong Provincial Key Laboratory of Artificial Intelligence in Medical Image Analysis and Application, Guangdong Provincial People's Hospital, Guangdong Academy of Medical Sciences, Guangzhou, 510080, China.}
\thanks{Hao Chen is with the Department of Computer Science and Engineering, The Hong Kong University of Science and Technology, Clear Water Bay, Hong Kong.}
\thanks{The first three authors contributed equally.}}

\begin{document}
\maketitle

\IEEEtitleabstractindextext{\begin{abstract}
Brain tumor segmentation (BTS) in magnetic resonance image (MRI) is crucial for brain tumor diagnosis, cancer management and research purposes. With the great success of the ten-year BraTS challenges as well as the advances of CNN and Transformer algorithms, a lot of outstanding BTS models have been proposed to tackle the difficulties of BTS in different technical aspects. However, existing studies hardly consider how to fuse the multi-modality images in a reasonable manner. In this paper, we leverage the clinical knowledge of how radiologists diagnose brain tumors from multiple MRI modalities and propose a clinical knowledge-driven brain tumor segmentation model, called CKD-TransBTS. Instead of directly concatenating all the modalities, we re-organize the input modalities by separating them into two groups according to the imaging principle of MRI. A dual-branch hybrid encoder with the proposed modality-correlated cross-attention block (MCCA) is designed to extract the multi-modality image features. The proposed model inherits the strengths from both Transformer and CNN with the local feature representation ability for precise lesion boundaries and long-range feature extraction for 3D volumetric images. To bridge the gap between Transformer and CNN features, we propose a Trans\&CNN Feature Calibration block~(TCFC) in the decoder. We compare the proposed model with five CNN-based models and six transformer-based models on the BraTS 2021 challenge dataset. Extensive experiments demonstrate that the proposed model achieves state-of-the-art brain tumor segmentation performance compared with all the competitors.
\end{abstract}
}

\maketitle
\IEEEdisplaynontitleabstractindextext

\IEEEpeerreviewmaketitle

\section{Introduction}
\label{sec:Introduction}
Glioma is the most common malignant tumor in the central nervous system~\cite{louis20212021}. Magnetic resonance imaging~(MRI) is the routine examination for glioma diagnosis. Conventional MRI, including pre- and post-contrast T1-weighted images (T1 and T1Gd), T2-weighted (T2) and T2-fluid-attenuated inversion recovery (T2FLAIR) images, provides valuable information for clinical diagnosis, therapeutic planning, and follow-up of gliomas~\cite{john2019multimodal}. Generally, radiologists integrate the diagnostic information across imaging modalities when assessing glioma, of which the enhancing regions, tumor necrosis, and peritumoral edema receive the most attention. For example, it is well accepted that the higher intensity of enhancement, larger area of necrosis and edema is associated with higher-grade gliomas, with worse prognosis. Therefore, automatically and precisely segmenting the lesion is a vital step for precision medicine in neurology, including treatment planning, quantitative analysis and research purposes.

Due to the strong feature representation capability, convolutional neural networks~(CNNs) have been widely used in the medical image segmentation tasks~\cite{hesamian2019deep} and achieved promising performance, including brain tumor segmentation (BTS)~\cite{liu2020deep}. Recently, vision transformer (ViT)~\cite{dosovitskiy2020image} brings the most powerful technique in natural language processing to the fields of computer vision and medical imaging~\cite{shamshad2022transformers}. Thanks to the self-attention mechanism, Transformer can capture long-range information, which perfectly fits 3D volumetric data. Thus, it has been rapidly adapted to brain tumor segmentation in 3D MRI sequences~\cite{hatamizadeh2022unetr, hatamizadeh2022unetformer}. Based on these two popular techniques, plenty of outstanding approaches have been proposed for brain tumor segmentation to tackle the following challenges, including lesion location and morphological uncertainty~\cite{wang2022relax}, low contrast~\cite{yu2020learning} and annotation bias~\cite{chen2021understanding}. However, the existing works overlook an important point on how to fuse the multi-modality images in a reasonable manner. Most of them fuse the modalities in either the input level or the feature level.

In recent AI research, multi-modal fusion has become one of the hottest topics~\cite{xu2022multimodal}. Many studies try to build the semantic connection between two different modalities, like image and language. However, brain MRI sequences are quite different from the other ones. In brain MRI images, there exists a very strong structural correlation between paired image modalities, which is the clue for brain tumor assessment. To be more specific, T1Gd is obtained based on the T1 with intravenous gadolinium contrast, and the enhancing regions indicate the disruption (or lack) of the blood-brain barrier, which is consistent with viable tumor and tumor-infiltrated brain. T2 and T2-FLAIR are often jointly interpreted. This clinical knowledge could be very useful in brain tumor segmentation.

Inspired by this, we propose a clinical knowledge-driven BTS model, named CKD-TransBTS. Instead of directly concatenating all the modalities, we simply re-organize the input modalities into two groups (T1 \& T1Gd) and (T2 \& T2FLAIR), according to their imaging principle. In the encoding phase, we design a dual-branch hybrid encoder with our proposed Modality-Correlated Cross-Attention block (MCCA) for multi-modal fusion and feature extraction. The hybrid encoder leverages the strengths of both Transformer and CNN. Transformer captures long-range information from adjacent slices in the 3D volumetric images. CNN introduces inductive bias for more precise lesion boundaries. In the decoding phase, we propose a Trans\&CNN Feature Calibration block (TCFC) in order to alleviate the bias of the features extracted from Transformer and CNN.

Extensive experiments are conducted to evaluate the effectiveness of the proposed CKD-TransBTS in the BraTS challenge 2021 dataset~\cite{baid2021rsna}. Our model outperforms five CNN-based models (including the 1st in the BraTS21 challenge validation phase) and six transformer-based models and achieved state-of-the-art BTS performance. Several ablation studies are introduced to validate the technical novelties, especially the clinical knowledge-driven formulation. The main contributions of this study are three-fold:
\begin{itemize}
	\item We propose a clinical knowledge-driven BTS model by considering the structural correlation between different image modalities and re-grouping the input images in a more reasonable manner.
	
	\item We propose two technical novelties in CKD-TransBTS. First, a dual-branch hybrid encoder with the novel Modality-Correlated Cross-Attention block (MCCA) is designed for multi-modal fusion and feature extraction. Second, a novel Trans\&CNN Feature Calibration block (TCFC) is proposed to bridge the gap and alleviate the bias of the features between Transformer and CNN.
	
	\item We have conducted a series of experiments on the BraTS21 dataset. Our proposed model achieves SOTA performance compared with five CNN-based models and six transformer-based models.
\end{itemize}

\section{Related Works}
\label{sec:Related Work}
Thanks to the great efforts of the Radiological Society of North America (RSNA), the American Society of Neuroradiology (ASNR), the Medical Image Computing and Computer Assisted Interventions (MICCAI) society and the BraTS challenge organizers~\cite{baid2021rsna,menze2014multimodal}, more and more standardized and well-labeled MR images have been released to promote the BTS algorithms. Currently, convolutional neural networks have dominated brain tumor segmentation~\cite{wadhwa2019review}. With the great success of the multi-head self-attention mechanism, transformer-based models are equally important in medical image segmentation~\cite{shamshad2022transformers}, especially for 3D volumetric images.

In this section, we first introduce the existing works in CNN-based BTS models and Transformer-based BTS models. Since we design a novel multi-modal fusion model, multi-modal fusion models are also reviewed in the third part.

\subsection{CNN-based BTS Models}
Recent CNN-based models have demonstrated promising performance in brain tumor segmentation~\cite{liu2020deep}. Restricted by the computational and memory resources, earlier CNN-based approaches~\cite{zhao2018deep,mehrtash2020confidence} segment the 2D MRI in a slice-by-slice manner. However, 2D approaches neglect the 3D sequential information.

Currently, more and more 3D BTS models are proposed to leverage 3D spatial information. nnU-Net~\cite{isensee2021nnu} is a general and adaptive baseline model for both 2D and 3D medical image segmentation, which derives a series of nnU-Net-based BTS models~\cite{luu2021extending,futrega2021optimized}. Liu~\etal~\cite{liu2021canet} propose CANet to capture the sequential information by introducing feature interaction graphs. Combining feature interaction graphs with convolutional space, CANet can capture the discriminative features with contexts. Zhou~\etal~\cite{zhou2020afpnet} enlarges the receptive field to capture the contextual information around the lesion and proposes lossless feature computation by employing the 3D atrous-convolution layer. They incorporate the multi-scale contexts and lesion information by an atrous-convolution feature pyramid for brain tumor segmentation. OM-Net~\cite{zhou2020one} integrates three correlated tasks into one network to achieve coarse-to-fine segmentation in a lightweight way and handles the different contributions of each channel for categories by using a CGA module. 

Since BTS involves 3D volumetric images from four MRI modalities, researchers now try to innovate the BTS models in the following two aspects. 1) How to leverage the 3D sequential information and the locality information. 2) How to fuse the multi-modal images.

\subsection{Transformer-based BTS Models}
Transformer~\cite{vaswani2017attention} is the most popular technology in the Natural Language Processing~(NLP) field thanks to the multi-head self-attention mechanism. ViT~\cite{khan2021transformers, han2022survey} extends transformer to computer vision by tokenizing the images.

Chen~\cite{chen2021transunet} takes advantage of both U-net structure and transformer and proposed a TransUNet for medical image segmentation. TransBTS~\cite{wang2021transbts} extracts global features by applying transformer blocks in the bottleneck layer. CoTr~\cite{xie2021cotr} captures the long-range dependency between encoder and decoder by introducing a deformable self-attention mechanism of the DeTrans-encoder. UNETR~~\cite{hatamizadeh2022unetr} learns contextual and long-range information by a Transformer-based encoder and fused localized information and global information in the skip connection. VT-UNet~\cite{peiris2021volumetric} simultaneously encodes local and global cues and captures fine detail for boundary refinement by the volumetric transformer encoder-decoder structure. nnFormer~\cite{zhou2021nnformer} interleaves convolution and self-attention operations to give full play to their strengths to eliminate the gap between features of encoder and decoder by using skip attention. Swin-UNETR~\cite{hatamizadeh2022swin} utilizes a Swin Transformer-based encoder to learn multi-scale contextual representations and model long-range dependencies.

The above studies provide excellent insights on how to associate transformer with CNN in brain tumor segmentation tasks. In this paper, we design a hybrid transformer model for BTS from different perspectives. 1) We formulate the clinical knowledge in the multi-modal fusion.
2) We introduce the inductive bias and the locality inside the transformer modal design, for a more harmonious mix of transformer and CNN. 3) We calibrate the features extracted from transformer and CNN, in order to bridge the gap between them.

\subsection{Multi-modal Fusion Models}\label{subsubsection:multi}
Multi-modal data provides richer information than a single modality does, which has attracted more and more attention in both natural data processing and medical data processing fields, such as visual question answering (VQA)~\cite{he2019exploring}, RGB-D object recognition~\cite{xu2017multi} and pathogenomics for prognosis analysis~\cite{ning2021multi}. For most of the above tasks, there exists a huge gap between two different modalities, such as \{text, image, speech\} and \{whole slide image, genomic data\}. These data share the same semantic features but different structural features. Different from that, MRI modalities in BTS are well aligned at pixel-level. It is hard to directly apply existing multi-modal fusion methods to BTS. Therefore, even though a lot of multi-modal fusion studies have been proposed, multi-modal fusion is still under-studied in BTS.

In early BTS models, most of them do not consider the multi-modal fusion problem. They just simply concatenate all the modalities before feeding them into the model. Now they try to fuse multiple modalities in a more reasonable manner. Zhang~\etal~\cite{zhang2020exploring} introduce a learnable weight to define the contribution of each modality. Wang~\etal~\cite{wang2020modality} design the model to learn the complementary information effectively by leveraging two same structural densely-connected branches to map two pairs of modalities. Zhou~\etal~\cite{zhou2020multi} introduce four independent branches for four modalities and fuse the features in the latent space. Zhang~\etal~\cite{zhang2021cross} learn the cross-modal feature representation by a GAN-based generation model. 

In this study, we introduce clinical knowledge of the imaging principles of different MRI sequences to guide the multi-modal fusion in a reasonable manner. By simply re-grouping the input images with a dual-branch hybrid encoder, the proposed model can learn better cross-modal feature representation.
\section{Methodology}\label{sec:Methodology}
\begin{figure*}
	\centering
	\includegraphics[width=.985\textwidth]{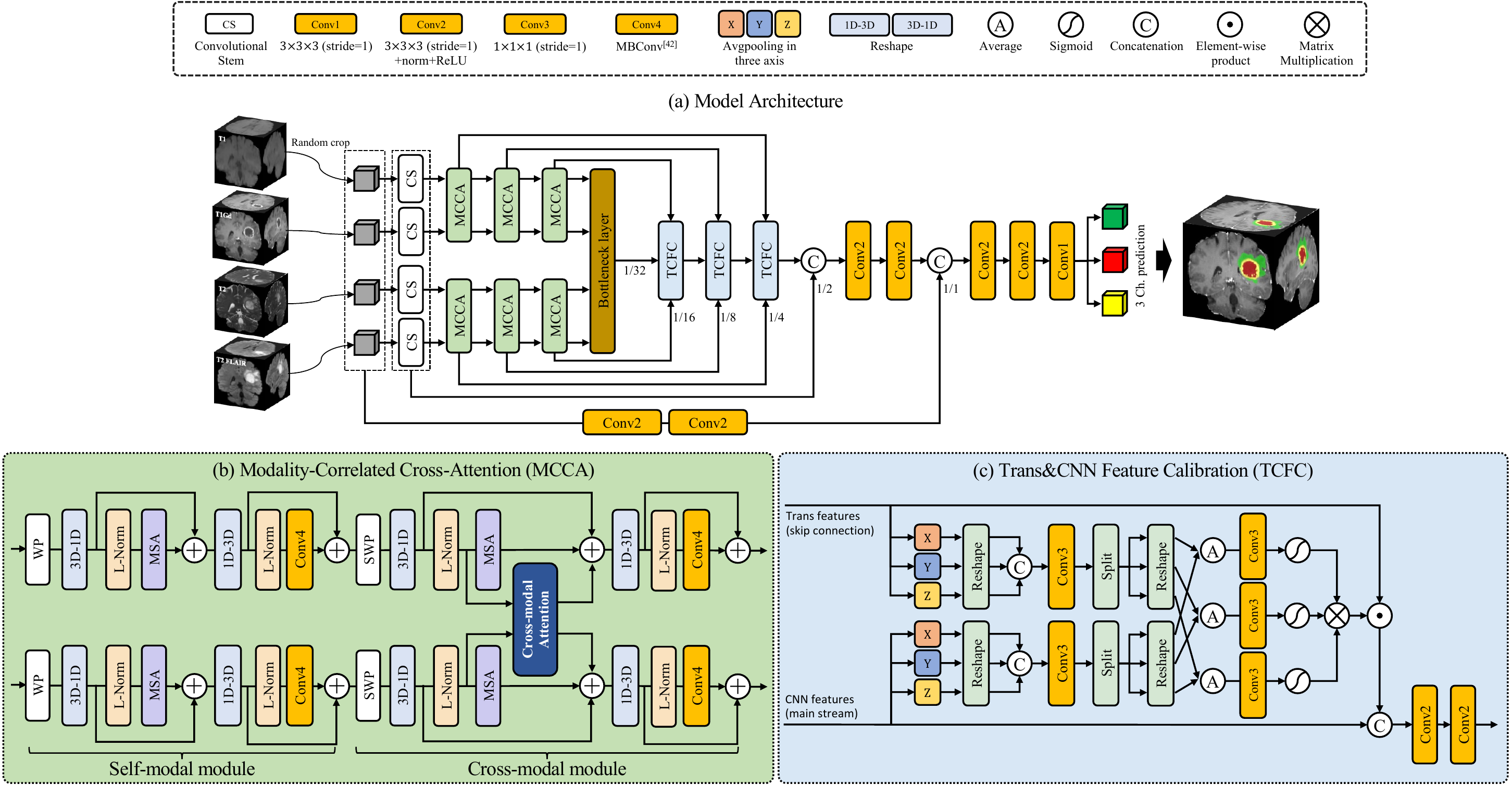}
	\caption{The architecture of the proposed \textbf{CKD-TransBTS}. (a) This model is a U-Net-like structure with a dual-branch hybrid encoder and a feature calibration decoder. According to the clinical knowledge of the MRI in brain tumor diagnosis, we separate the input images into two groups (T1 \& T1Gd) and (T2 \& T2FLAIR). Convolutional stem is introduced at the beginning. The encoder comprises several MCCA blocks ((b) Modality-Correlated Cross-Attention) which enables cross-modal interactions in a reasonable manner. The decoder consists of several TCFC blocks ((c) Trans\&CNN Feature Calibration) to bridge the semantic gap between the features extracted by transformer and CNN. After several convolutional blocks, the model predicts the final brain tumor segmentation results. Note that, in the encoding (decoding) phase, we downsample (upsample) the feature maps by a convolutional (deconvolutional) layer at the end of each stage. In this figure, we omit the downsample and upsample operations for simplification. The resolutions of the feature maps are specified at each stage by the scaling factors.}\label{fig:overall}
\end{figure*}

In this section, we first demonstrate the insight of the clinical knowledge-driven formulation and the overall architecture of our model in Sec.~\ref{subsection:formulation}. Then, we introduce the knowledge-based dual-branch hybrid encoder with the proposed Modality-Correlated Cross-Attention Module~(MCCA) block in Sec.~\ref{subsection:encoder}. The details of the feature calibration decoder are given in Sec.~\ref{subsection:decoder} with Trans\&CNN Feature Calibration~(TCFC) module. Sec.~\ref{sec:implement} shows the details of implementation in this study.

\subsection{Formulation and Model Architecture}~\label{subsection:formulation}
Before going into the details of the model, let us start by answering the question of `\textit{how radiologists diagnose brain tumor?}', which inspires and motivates the formulation of our proposed model.
\subsubsection{How Radiologists Diagnose Brain Tumor?}
MRI is the routine clinical examination for brain tumor diagnosis. It usually contains four imaging sequences (modalities), including T1-weighted, T1Gd, T2-weighted and T2FLAIR. Generally, radiologists integrate the diagnostic information across imaging modalities when assessing brain tumors. T1-weighted is the pre-contrast sequence which is the basic imaging modality to pre-locate the brain tumor. T1Gd is the post-contrast sequence with the infused gadolinium enhancing the vascular structures and whether the blood-brain barrier is broken down. Therefore, T1-weighted and T1Gd are usually paired to define the tumor core. T2-weighted imaging is used to detect the free water. In brain tumors, T2-weighted and T2FLAIR images are often jointly interpreted. For the T2-weighted hyperintense, non-enhancing regions in glioblastoma, those that contain free water (e.g. tumor necrosis) frequently present with T2FLAIR hypointensity, while those contain bound water (e.g. vasogenic edema) appear as hyperintense T2FLAIR signal~\cite{john2019multimodal}.

\subsubsection{Clinical Knowledge-Driven Formulation}
Inspired by the way that radiologists assess brain tumors in MRI images, we want the model to learn the spatial and structural correlation between two correlated sequences. Given four image modalities $\{\mathcal{X}_{\rm T1},\mathcal{X}_{\rm T1Gd}, \mathcal{X}_{\rm T2},\mathcal{X}_{\rm T2FLAIR}\}$ and the segmentation model $f$ with the model parameter $\theta$. Most of the existing BTS models simply concatenate all the input modalities and feed them into the segmentation model at once to predict the segmentation result $\mathcal{S}$.
\begin{equation}
\mathcal{S}=f(\theta,\{\mathcal{X}_{\rm T1},\mathcal{X}_{\rm T1Gd}, \mathcal{X}_{\rm T2},\mathcal{X}_{\rm T2FLAIR}\})
\end{equation}

In this study, we re-organize the order of the input images by grouping each two correlated imaging modalities $\{\mathcal{X}_{\rm T1},\mathcal{X}_{\rm T1Gd}\}$ and $\{\mathcal{X}_{\rm T2},\mathcal{X}_{\rm T2FLAIR}\}$, according to the clinical knowledge.
\begin{equation}
\mathcal{S}=f_{\rm ours}(\theta,(\{\mathcal{X}_{\rm T1},\mathcal{X}_{\rm T1Gd}\}, \{\mathcal{X}_{\rm T2},\mathcal{X}_{\rm T2FLAIR}\}))
\end{equation}

By grouping two correlated image modalities together, our model can learn the inherent correlation between two image modalities, resulting in a better cross-modal feature representation.

\subsubsection{Model Architecture}
Fig.~\ref{fig:overall}~(a) demonstrates the model architecture of the proposed CKD-TransBTS. The same with most of the segmentation models, CKD-TransBTS keeps the U-Net~\cite{ronneberger2015u} like structure with skip connections. Since MRI images are 3D volumetric data, we use swim transformer~\cite{liu2021swin} as the basic architecture of the proposed model to capture the long-range information. In order to bring inductive bias and encourage better local feature representation, we associate transformer with CNN by introducing convolutional layers inside the transformer model.

Since there are two groups of input images, we design a dual-branch hybrid encoder with a convolutional stem and several MCCA blocks. The MCCA block exchanges the information between two correlated image modalities by the cross-modal attention. All the multi-modal features are finally fused in a bottleneck layer. A feature calibration decoder with several TCFC blocks and convolutional layers is designed to obtain the final segmentation results.
\subsection{Dual-Branch Hybrid Encoder}~\label{subsection:encoder}
As shown in Fig.~\ref{fig:overall}, a dual-branch hybrid encoder is designed to extract the features from two groups of image modalities. Since two branches are identical (but do not share weights), we only show one branch with the image pair $\{\mathcal{X}_{\rm T1},\mathcal{X}_{\rm T1Gd}\}$ as the example for simplification. The other image pair $\{\mathcal{X}_{\rm T2},\mathcal{X}_{\rm T2FLAIR}\}$ is processed in the same way by the other branch.

\begin{figure}
	\centering
	\includegraphics[width=.8\linewidth]{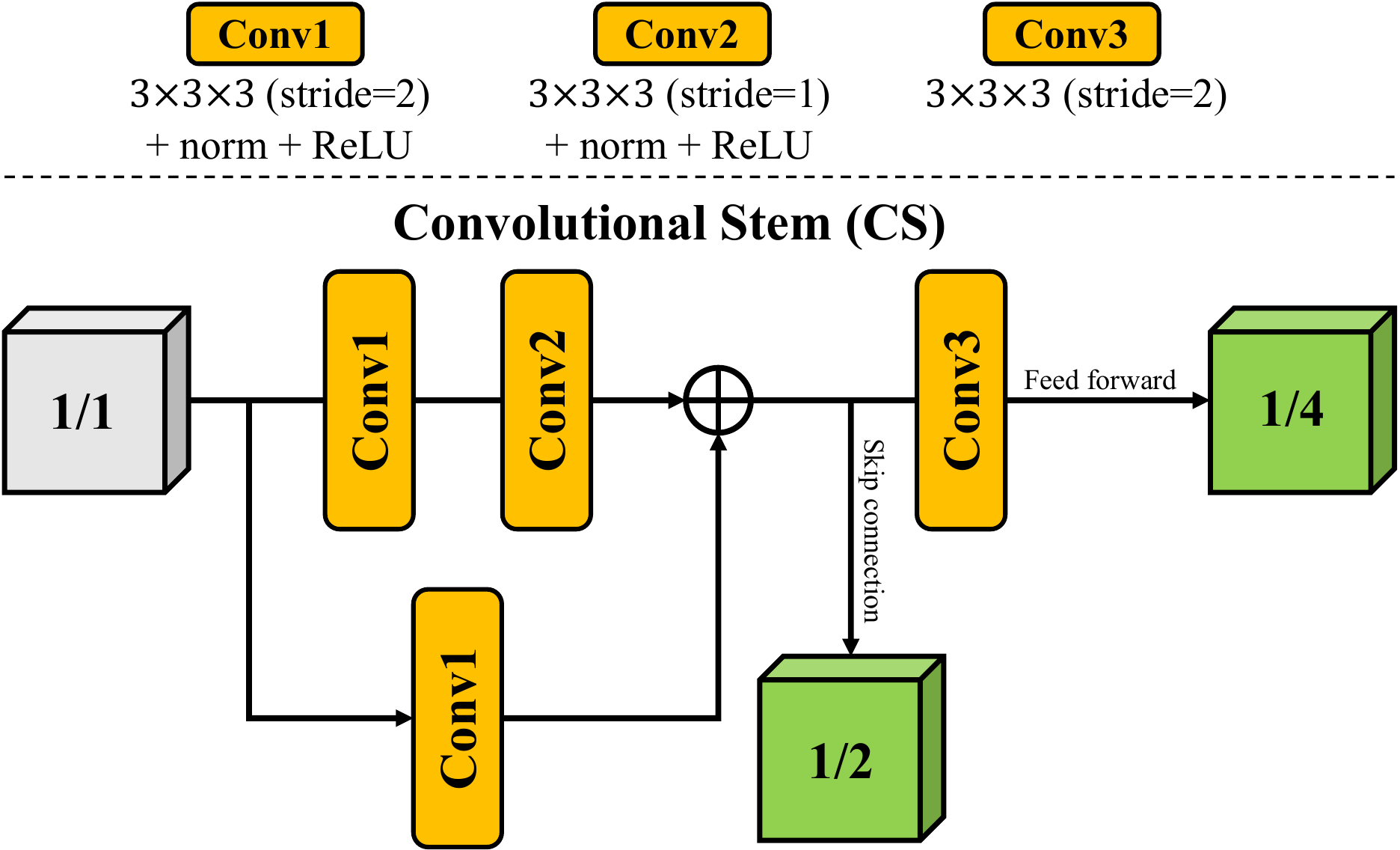}
	\caption{Convolution stem comprises several convolutional layers with different configurations to downsample the input image in a softer way. In this figure, the gray and green volumes represent the original input image and the intermediate outputs, respectively. $\{1/1, 1/2, 1/4\}$ indicate the scaling factors of the volumes.}
	\label{fig:conv_stem}
\end{figure}

\subsubsection{Convolutional Stem~(CS)}
Downsampling is a common way to reduce the input dimension to save computational and memory resources when processing MRI sequences. However, conventional image downsampling approaches, such as nearest neighbor or bilinear interpolation, will cause information loss. And for the BTS task with four volumetric input images, existing approaches tend to downsample the input images by four. In order to reduce the input dimension in a softer way, we introduce the convolutional stem (CS)~\cite{xiao2021early} for each image modality.

As shown in Fig.~\ref{fig:conv_stem}, CS comprises several convolutional blocks with different configurations. Given the input image $\mathcal{X}_{\rm T1}$, CS outputs two feature volumes $\mathcal{C}_{\rm T1}^{\frac{1}{2}}$ and $\mathcal{C}_{\rm T1}^{\frac{1}{4}}$ under two different scales step-by-step, shown as follows.
\begin{equation}\label{eq:stem}
\{\mathcal{C}_{\rm T1}^{\frac{1}{2}}, \mathcal{C}_{\rm T1}^{\frac{1}{4}} \}= \mathbf{CS}(\mathcal{X}_{\rm T1})
\end{equation}
where $\mathcal{C}_{\rm T1}^{\frac{1}{2}}$ is the feature volume for skip connection and $\mathcal{C}_{\rm T1}^{\frac{1}{4}}$ is the one for feed forward to the MCCA block.

There are two advantages of CS in the BTS task. First, compared with directly downsampling input images by four, CS provides two different scales of feature volumes to help recover the information in the decoding phase. The second advantage originates from Xiao~\etal~\cite{xiao2021early} that early convolution operation can increase the optimization stability for ViT.

\subsubsection{Modality-Correlated Cross-Attention~(MCCA) Block}\label{sec:MCCA}
We design a modality-correlated cross-attention~(MCCA) block to extract the cross-modal features. The MCCA block requires a paired inputs and generates a paired outputs. 
To consider the locality and the long-range information simultaneously, we combine transformer and CNN in the MCCA block. As shown in Fig.~\ref{fig:overall}~(b), MCCA block consists of two identical branches which are used to extract features from two modalities individually. Each branch is composed of two cascaded modules, the self-modal module and the cross-modal module.

\textbf{Self-modal module:}~The self-modal module serves for feature extraction of each single modality, which is a hybrid Transformer-CNN module. We first use the transformer to capture the long-range information. And then we introduce inductive bias and locality by replacing the MLP layer by the convolution layers.
\begin{equation}\label{eq:self-modal1}
\left\{
\begin{aligned}
&\mathcal{F}_\mathrm{T1}^{l} = \mathrm{MSA}(\mathrm{LN}(\mathcal{F}_\mathrm{T1}^{l-1}))+\mathcal{F}_\mathrm{T1}^{l-1} \\
&\mathcal{F}_\mathrm{T1}^{l+1} = \mathrm{MBConv}(\mathrm{LN}(\hat{\mathcal{F}}_\mathrm{T1}^{l}))+\mathcal{F}_\mathrm{T1}^{l}\\
\end{aligned}
\right.
\end{equation}

\begin{equation}\label{eq:self-modal2}
\left\{
\begin{aligned}
&\mathcal{F}_\mathrm{T1Gd}^{l} = \mathrm{MSA}(\mathrm{LN}(\mathcal{F}_\mathrm{T1Gd}^{l-1}))+\mathcal{F}_\mathrm{T1Gd}^{l-1} \\
&\mathcal{F}_\mathrm{T1Gd}^{l+1} = \mathrm{MBConv}(\mathrm{LN}(\mathcal{F}_\mathrm{T1Gd}^{l}))+\mathcal{F}_\mathrm{T1Gd}^{l}
\end{aligned}
\right.  
\end{equation}
For simplicity, we omit the reshape operations in Eq.~\ref{eq:self-modal1} and Eq.~\ref{eq:self-modal2}. $\mathrm{MSA}(\cdot)$ is the multi-head self-attention~(MSA) with window partition. $\mathrm{LN}(\cdot)$ represents the layer normalization and $\mathrm{MBConv}(\cdot)$ is from EfficientNet~\cite{tan2021efficientnetv2}.

\textbf{Cross-modal module:}
The cross-modal module follows the swim transformer~\cite{liu2021swin} with shifted window partition and also replaces MLP layer by the MBConv. It exchanges the information between two correlated modalities by a cross-modal attention $\mathrm{CM\textrm{-}MSA}(\cdot)$, defined as follows.
\begin{equation}
\mathcal{M}_\mathrm{T1}, \mathcal{M}_\mathrm{T1Gd} = \mathrm{CM\textrm{-}MSA}(\mathrm{LN}(\mathcal{F}_\mathrm{T1}^{l+1}),\mathrm{LN}(\mathcal{F}_\mathrm{T1Gd}^{l+1})))
\end{equation}
\begin{equation}
\mathcal{M}_\mathrm{T1}=\mathrm{SoftMax}(\frac{Q_\mathrm{T1}K_\mathrm{T1Gd}^T}{\sqrt{d}}+B)V_\mathrm{T1Gd}
\end{equation}
\begin{equation}
\mathcal{M}_\mathrm{T1Gd}=\mathrm{SoftMax}(\frac{Q_\mathrm{T1Gd}K_\mathrm{T1}^T}{\sqrt{d}}+B)V_\mathrm{T1}
\end{equation}

The design of cross-modal module is shown as follows.
\begin{equation}\label{eq:cross-modal1}
\left\{
\begin{aligned}
&\mathcal{F}_\mathrm{T1}^{l+2} = \mathrm{MSA}(\mathrm{LN}(\mathcal{F}_\mathrm{T1}^{l+1}))+ \mathcal{M}_\mathrm{T1} + \mathcal{F}_\mathrm{T1}^{l+1} \\
&\mathcal{F}_\mathrm{T1}^{l+3} = \mathrm{MBConv}(\mathrm{LN}(\mathcal{F}_\mathrm{T1}^{l+2}))+\mathcal{F}_\mathrm{T1}^{l+2}
\end{aligned}
\right.  
\end{equation}
\begin{equation}\label{eq:cross-modal2}
\left\{
\begin{aligned}
&\mathcal{F}_\mathrm{T1Gd}^{l+2} = \mathrm{MSA}(\mathrm{LN}(\mathcal{F}_\mathrm{T1Gd}^{l+1}))+ \mathcal{M}_\mathrm{T1Gd} + \mathcal{F}_\mathrm{T1Gd}^{l+1} \\
&\mathcal{F}_\mathrm{T1Gd}^{l+3} = \mathrm{MBConv}(\mathrm{LN}(\mathcal{F}_\mathrm{T1Gd}^{l+2}))+\mathcal{F}_\mathrm{T1Gd}^{l+2}
\end{aligned}
\right.  
\end{equation}
where $\mathcal{F}_\mathrm{T1}^{l+3}$ and $\mathcal{F}_\mathrm{T1Gd}^{l+3}$ are the outputs of the MCCA block.

\subsubsection{Bottleneck Layer}
After three MCCA blocks, we concatenate the features of four modalities and feed them into a bottleneck layer which is introduced to bridge the encoder and the decoder. The bottleneck layer shares the same structure with the single branch of the MCCA block without cross-modal attention. The output of the bottleneck layer is defined as $\mathcal{F}_\mathrm{BNL}$.

\subsection{Feature Calibration Decoder}~\label{subsection:decoder}
In this part, we design a feature calibration decoder to predict the final segmentation results. As shown in Fig.~\ref{fig:overall}, the intermediate features extracted by the encoder are passed to the decoder by skip connections. Since the encoder is a hybrid model which is composed of both transformer and CNN. And the decoder is a pure CNN-based design. There exist a semantic gap between the features of the encoder and the decoder. To bridge the gap, we propose a Trans\&CNN Feature Calibration block (TCFC). The feature calibration decoder contains three consecutive TCFC blocks, several convolutional blocks and a segmentation head.


\subsubsection{Trans\&CNN Feature Calibration Block (TCFC)}
Fig.~\ref{fig:overall}~(c) demonstrates the architecture of the TCFC block whose purpose is to bridge the semantic gap between MCCA and TCFC blocks by providing pixel-wise spatial attention to the features extracted by the MCCA block. Let us denote the feed-forward feature tensor, the transformer feature tensor from skip connections and the output feature tensor as $\mathcal{F}$, $\mathcal{F}_\mathrm{trans}$ and $\mathcal{F}'$, TCFC block is formulated by Eq.~\ref{eq:TCFC}.
\begin{equation}\label{eq:TCFC}
\begin{aligned}
\mathcal{F}'=&\mathrm{TCFC}(\mathcal{F}_\mathrm{trans},\mathcal{F})\\
\mathcal{F}_\mathrm{trans}=&\mathrm{Concate}(\mathcal{F}_\mathrm{trans}^{b1},\mathcal{F}_\mathrm{trans}^{b2})
\end{aligned}
\end{equation}
where $\mathcal{F}_\mathrm{trans}^{b1}$ and $\mathcal{F}_\mathrm{trans}^{b2}$ denote the tensors from two branches in the dual-branch hybrid encoder. In the first TCFC block, $\mathcal{F}=\mathcal{F}_\mathrm{BNL}$.

Since the input and the feature tensors are 3D volume. In order to make full use of the 3D information, average pooling is applied in three directions separately for both $\mathcal{F}$ and $\mathcal{F}_\mathrm{trans}$.
\begin{equation}
\left\{
\begin{aligned}
&\mathcal{F}^X = \mathrm{Avepool(\mathcal{F})}, \mathcal{F}^X \in \mathbb{R}^{c\times x\times1 \times 1}\\
&\mathcal{F}^Y = \mathrm{Avepool(\mathcal{F})}, \mathcal{F}^Y \in \mathbb{R}^{c\times 1\times y \times 1}\\
&\mathcal{F}^Z = \mathrm{Avepool(\mathcal{F})}, \mathcal{F}^Z \in \mathbb{R}^{c\times 1\times1 \times z}\\
\end{aligned}
\right.
\end{equation}

\begin{equation}
\left\{
\begin{aligned}
&\mathcal{F}^X_\mathrm{trans} = \mathrm{Avepool(\mathcal{F}_\mathrm{trans})}, \mathcal{F}^X_\mathrm{trans} \in \mathbb{R}^{c\times x\times1 \times 1 }\\
&\mathcal{F}^Y_\mathrm{trans} = \mathrm{Avepool(\mathcal{F}_\mathrm{trans})}, \mathcal{F}^Y_\mathrm{trans} \in \mathbb{R}^{c\times 1\times y \times 1}\\
&\mathcal{F}^Z_\mathrm{trans} = \mathrm{Avepool(\mathcal{F}_\mathrm{trans})}, \mathcal{F}^Z_\mathrm{trans} \in \mathbb{R}^{c\times 1\times1 \times z}\\
\end{aligned}
\right.
\end{equation}
where $x=y=z$ since the input images are cubes in this study.

By reshaping vectors of three directions into the same shape, we concatenate them and compress the channels by a $1\times 1 \times 1$ convolutional layer. Then we can obtain new feature vectors for three directions $\{\hat{\mathcal{F}}^X_\mathrm{trans},\hat{\mathcal{F}}^Y_\mathrm{trans},\hat{\mathcal{F}}^Z_\mathrm{trans}\}$ and $\{\hat{\mathcal{F}}^X,\hat{\mathcal{F}}^Y,\hat{\mathcal{F}}^Z\}$ by splitting them back to the original dimension. 
Next, we aggregate the features from transformer and CNN in a direction-wise manner.
\begin{equation}
\left\{
\begin{aligned}
\bar{\mathcal{F}}^X=\mathrm{Sigmoid}(\mathrm{Conv}(\mathrm{Avg}(\hat{\mathcal{F}}^X_\mathrm{trans},\hat{\mathcal{F}}^X)))\\
\bar{\mathcal{F}}^Y=\mathrm{Sigmoid}(\mathrm{Conv}(\mathrm{Avg}(\hat{\mathcal{F}}^Y_\mathrm{trans},\hat{\mathcal{F}}^Y)))\\
\bar{\mathcal{F}}^Z=\mathrm{Sigmoid}(\mathrm{Conv}(\mathrm{Avg}(\hat{\mathcal{F}}^Z_\mathrm{trans},\hat{\mathcal{F}}^Z)))
\end{aligned}
\right.
\end{equation}

By a matrix multiplication operation of three vectors, we can obtain a calibrated attention tensor $A$.
\begin{equation}
\mathcal{A} =\bar{\mathcal{F}}^X \bar{\mathcal{F}}^Y \bar{\mathcal{F}}^Z
\end{equation}
Then the output of TCFC block can be obtained by concatenating the calibrated transformer features and the feed forward main stream features.
\begin{equation}
\mathcal{F}'= \mathrm{Concate}(\mathcal{A}\mathcal{F}_\mathrm{trans},\mathcal{F})
\end{equation}

\subsection{Implementation Details}\label{sec:implement}
We implement all the experiments on the PyTorch and MONAI, and train the models on a workstation with an NVIDIA 3090 GPU. We set the learning rate to $1e-4$ and adjust it using the cosine annealing algorithm~\cite{loshchilov2016sgdr}. The number of epochs for model training is 500 and the Dice loss is used as the objective function. For each modality, the initial sub-volume resolution is $4\times4\times4$ and the initial embedding size is 32. In the training phase,  we first obtain the minimum bounding box of the volume and then partition it randomly into a volume size of 128$\times$128$\times$128. To make the data distribution more complex and alleviate the over-fitting problem, we apply several data augmentation methods, including random zoom, random flip in three directions, Gaussian noise, Gaussian blur and random contrast. In the test phase, we use the sliding window method with an overlap rate of 0.6. 

\begin{table*}[t]
	\centering
	\caption{Quantitative comparison with SOTA methods in BraTS21 dataset ($\dagger$ means CNN-based models). The top-3 results are in \red{red}, \blue{blue} and \green{green}.}
	\label{tab:quantitative}
	\begin{tabular}{l|c|c|cccc|cccc}
		\hline 
		\multirow{2}{*}{Models}&\multirow{2}{*}{Publication}&\multirow{2}{*}{Year}&\multicolumn{4}{c|}{Dice $\uparrow$} &\multicolumn{4}{c}{HD95 (mm) $\downarrow$} \\
		&& &ET     &TC   &WT      &Mean    &ET     &TC     &WT  &Mean\\\hline
		$\dagger$VNet~\cite{milletari2016v}  &3DV  &2016    &0.7820  &0.8051  &0.8402  &0.8091 &20.80  &25.08  &15.69 &20.52\\
		$\dagger$ResUNet~\cite{zhang2018road} &IEEE GRSL  &2017  &0.8143  &0.8472  &0.9027  &0.8547  &14.30  &9.33  &10.12 &11.25 \\
		$\dagger$UNet++~\cite{zhou2019unet++}  &TMI  &2019    &0.7813  &0.8225  &0.8493  &0.8177 &23.88  &17.04  &8.64 &16.52\\
		$\dagger$AttentionUNet~\cite{schlemper2019attention}  &MIA  &2019    &0.7897  &0.8373  &0.8566  &0.8278 &18.47  &11.84  &15.97 &15.43\\
		
		$\dagger$DynUNet (1st BraTS21)~\cite{futrega2021optimized} &arXiv  &2021     &0.8581  &0.8971  &\blue{0.9288}  &0.8946 &13.03 &\green{6.71}  &\green{6.99} &8.91 \\\hline

		TransBTS~\cite{wang2021transbts} &MICCAI&2021   &0.8181  &0.8500  &0.8795 &0.8494 &16.79  &11.14  &12.78 &13.57 \\
		TransUNet~\cite{chen2021transunet}   &arXiv&2021  &0.8182  &0.8772  &0.9191 &0.8715 &13.09  &7.34  &\red{6.16} &8.86 \\
		VTNet~\cite{peiris2021volumetric} &arXiv&2021  &0.8551  &0.8822  &0.9134 &0.8837 &\blue{9.01}  &6.92  &8.22  &\blue{8.05}\\
		UNETR~\cite{hatamizadeh2022unetr} &WACV& 2022      &0.8520  &0.8664  &0.9220  &0.8803 &12.26  &7.73  &7.78  &9.26\\
		SegTransVAE~\cite{pham2022segtransvae}  &ISBI&2022     &\green{0.8622}  &\blue{0.8999}  &0.9254 &\green{0.8958} &\green{10.59}  &\red{5.88}  &7.71 &\green{8.06} \\
		Swin UNETR~\cite{hatamizadeh2022swin}   &arXiv&2022    &\blue{0.8681}  &\green{0.8998}  &\green{0.9273} &\blue{0.8984} &11.09  &6.89  &7.33 &8.44 \\\hline
		
		Ours &-&2022 &\red{0.8850} &\red{0.9016} &\red{0.9333} &\red{0.9066} &\red{5.93}  &\blue{6.54}  &\blue{6.20} &\red{6.22} \\
		\hline
	\end{tabular}
\end{table*}

\section{Experiment}
\label{sec:Experiment}
In this section, we first describe the dataset and the evaluation metrics in Sec.~\ref{sec:dataset}. Then, we compare our proposed model with existing SOTA approaches in Sec.~\ref{sec:sota}. In Sec.~\ref{sec:ablation}, we conduct ablation studies to evaluate the effectiveness of each technical novelty.

\subsection{Dataset and Evaluation Metrics}\label{sec:dataset}

\textbf{BraTS Challenge 2021:} The dataset provides a large amount of annotated brain tumor MRI data, mainly from The Cancer Imaging Archive~\cite{baid2021rsna,clark2013cancer}. Since the validation and test data in BraTS challenge 2021 is private, we split the training set (1251 3D MRI images) provided by the challenge organizers for all the experiments. The training, validation and test set contains 834, 208 and 209 samples respectively. All the MRI images were skull stripped and resampled to $1 mm^{3}$. Each patient's MRI includes four modalities: T1-weighted~(T1), post-contrast T1-weighted~(T1Gd), T2-weighted~(T2), and T2 Fluid Attenuated Inversion Recovery~(T2FLAIR), with co-registered to T1 anatomical template. Annotation is divided into three sub-regions: Gd-enhancing tumor~(ET), the peritumoral
edematous/invaded tissue~(ED) and the necrotic tumor core~(NCR). Suggested by the challenge organizers, these sub-regions can be clustered into three more segmentation-friendly regions which are used to evaluate the segmentation performance, including enhanced tumor~(ET), tumor core~(TC)~(joining ET and NCR), and whole tumor~(WT)~(joining ED to TC).

\begin{figure*}[t]
	\centering
	\includegraphics[width=0.99\linewidth]{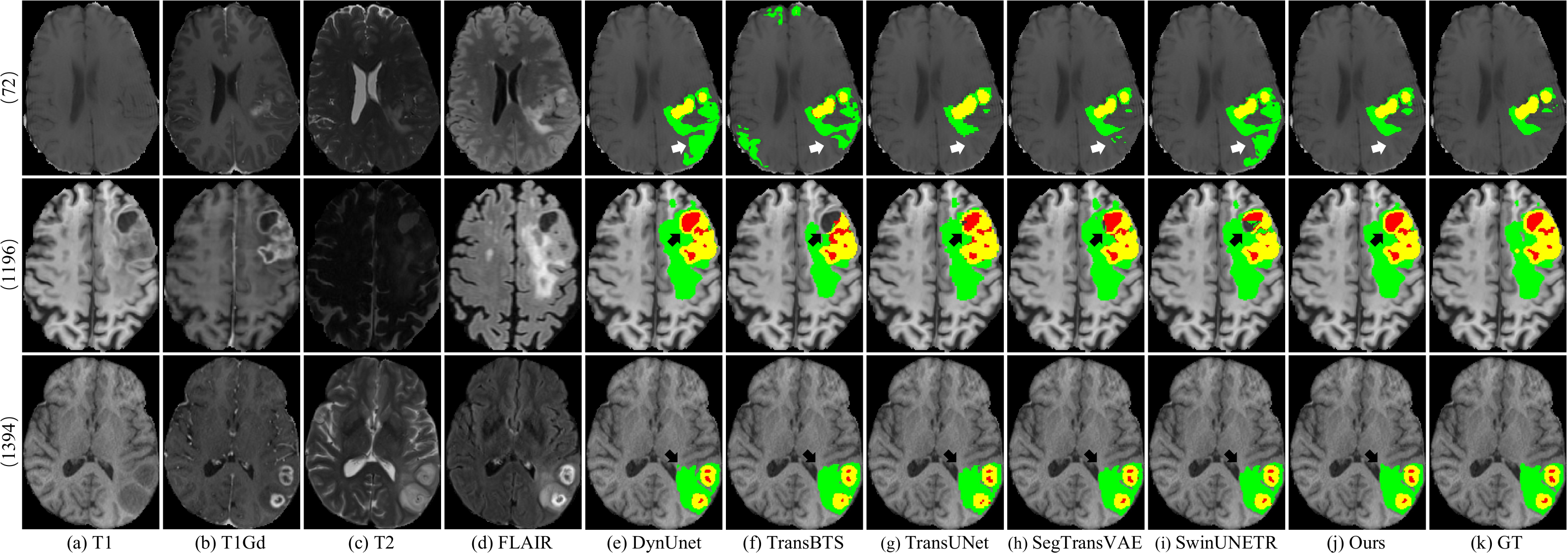}
	\caption{Visualization of quantitative comparison of SOTA methods on the BraTS21 dataset. From top to bottom are BraTS\_72, BraTS\_1196 and BraTS\_1394 respectively. Green, yellow and red regions indicate ED, ET and NCR. White arrows highlight some false positive results in the results of the baseline models. Black arrows demonstrate the superior regions of our results compared with the baseline models.}
	\label{fig:qualitative}
\end{figure*}

\begin{figure}[t]
	\centering
	\setlength{\tabcolsep}{1pt}
	\begin{tabular}{cccc}
		\includegraphics[width=0.245\linewidth]{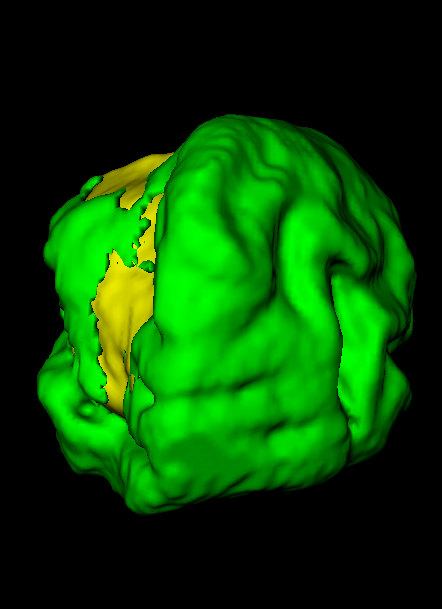}&
		\includegraphics[width=0.245\linewidth]{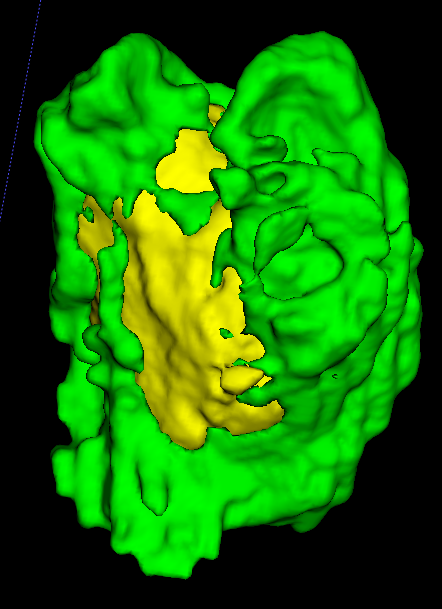}&
		\includegraphics[width=0.245\linewidth]{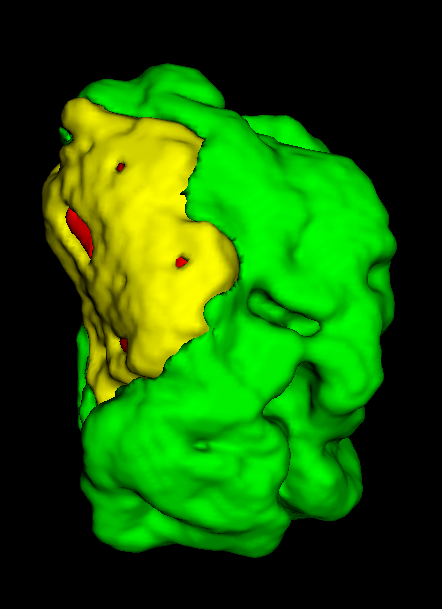}&
		\includegraphics[width=0.245\linewidth]{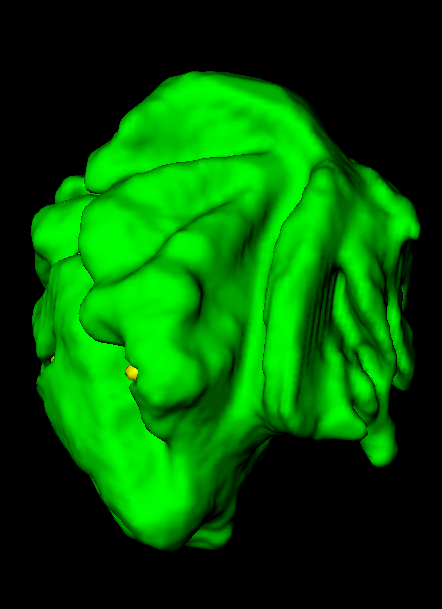}\\
		\includegraphics[width=0.245\linewidth]{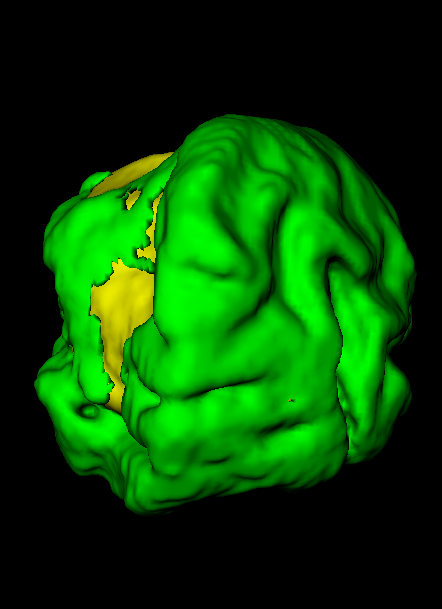}&
		\includegraphics[width=0.245\linewidth]{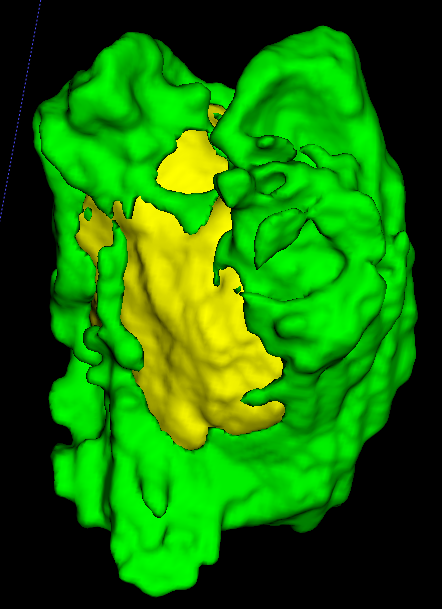}&
		\includegraphics[width=0.245\linewidth]{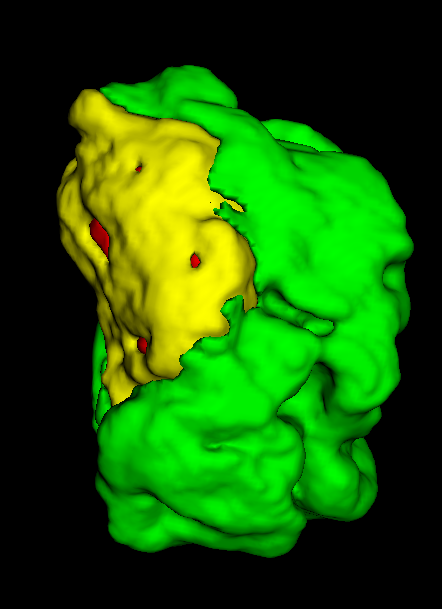}&
		\includegraphics[width=0.245\linewidth]{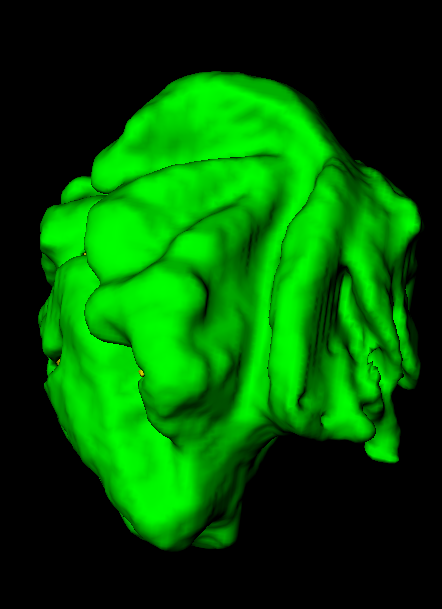}\\
		(a) 156 & (b) 416 & (c) 809 & (d) 1401
	\end{tabular}
	\caption{3D visualization of our s tumor segmentation results. The first row shows our segmentation results. The second row demonstrates the ground truth. The numbers indicate the indices of the samples. }
	\label{fig:3D-results}
\end{figure}

\textbf{Evaluation metrics:} In our experiments, we use Dice score and 95\% Hausdorff distance~(HD95) to evaluate the segmentation results.

\subsection{Comparisons with SOTA Models}\label{sec:sota}
We compare our proposed CKD-TransBTS model with several SOTA models, including five CNN-based models ( VNet~\cite{milletari2016v},
ResUNet~\cite{zhang2018road}, UNet++\cite{zhou2019unet++} AttentionUNet~\cite{schlemper2019attention} and DynUNet~\cite{futrega2021optimized}) and six transformer-based models (TransBTS~\cite{wang2021transbts}, TransUNet~\cite{chen2021transunet}, UNETR~\cite{hatamizadeh2022unetr}, VTNet~\cite{peiris2021volumetric}, SegTransVAE~\cite{pham2022segtransvae} and Swin UNETR~\cite{hatamizadeh2022swin}). For each baseline model in this experiment, we directly run the code if it has been released. For the baseline models without code, we implement them exactly following the details in the corresponding papers. All the baseline models and our proposed model are trained under the same computer hardware with the same dataset split. The best models are selected by the validation set. To be fair, all the quantitative and qualitative results are the direct output of the model without any post-processing procedure.

Table~\ref{tab:quantitative} demonstrates the quantitative results of all the models. The upper part and lower part demonstrate the CNN-based models (marked by $\dagger$) and the transformer-based models respectively. We color-code the top-3 approaches for every score in red, blue and green. As demonstrated in Table~\ref{tab:quantitative}, our proposed model outperforms all the CNN-based models and transformer-based models and achieves SOTA segmentation performance. Specifically, the Dice scores of the proposed CKD-TransBTS for ET, TC and WT all outperform the competitors. The HD95 metric for ET achieves the best (5.93 mm) which is 3 mm lower than the second one (9.01 mm). The HD95 metrics for TC and WT still maintain the top-2 performance with less than 1 mm performance gap. DynUNet, the winner of BraTS challenge 2021, achieves the best performance among the CNN-based models. However, Dice score and HD95 distance of ET are less effective than our proposed model (Dice: 0.8581 v.s. 0.8850, HD95: 13.03 v.s. 5.93). It proves the superiority of the clinical knowledge-driven formulation of our proposed model. Since the enhanced tumor (ET) is defined by comparing the difference of the hyperintense regions between pre- and post- contrast images (T1 and T1Gd). By simply re-grouping the input images, our model can achieve much better segmentation results in ET compared with other baseline models. The same observation can also be found when compared with transformer-based models. 

Fig.~\ref{fig:qualitative} demonstrates the qualitative results compared with several representative baseline models. As can be seen in the first row in Fig.~\ref{fig:qualitative}, our model successfully suppresses the false positive results compared with other baseline models (marked by the white arrows). In the second row, our model generates more precise segmentation results locally with fewer noises compared with all the other competitors, pointed by the black arrows. Therefore, it results in a lower HD95 distance. In the third row, our model achieves more complete and precise boundaries. Thanks to the clinical knowledge-driven design, CKD-TransBTS can aware of the correlations between the correlated modalities, to ensure spatial and contextual correctness. Since the clinical knowledge-driven idea originates from how radiologists diagnose brain tumors, this formulation also makes the results closer to the radiologists. The hybrid model of transformer and CNN inherits the strengths from both of them. Capturing long-range information can avoid some false positive results far from the lesions. Inductive bias and locality can achieve locally more precise segmentation results. We also demonstrate the 3D volumetric segmentation results compared with ground truth. Fig.~\ref{fig:3D-results} shows that our model can generate 3D segmentation results very close to the ground truth.

\subsection{Ablation Studies}\label{sec:ablation}
We also conduct several ablation studies to evaluate the superiority of each technical novelty in our proposed model, including the clinical knowledge-driven multi-modal fusion, the hybrid encoder of transformer and CNN, and the feature calibration decoder. We compare our final model with several models with different configurations. (1) The baseline model without cross-modal fusion, the hybrid encoder and feature calibration. (2)-(4) are the baseline models with only one technical novelty. (5)-(7) are the baseline models with two technical novelties. (8) is our final model. The quantitative results are illustrated in Table~\ref{tab:ablation}.


\begin{table*}[t]
	\centering
	\caption{Ablation studies on multi-modal fusion, hybrid encoder and feature calibration.}
	\label{tab:ablation}
	\begin{tabular}{c|ccc|cccc|cccc}
		\hline 
		\multirow{2}{*}{Models}&\multicolumn{3}{c|}{Ablation} &\multicolumn{4}{c|}{Dice $\uparrow$} & \multicolumn{4}{c}{HD95 $\downarrow$} \\
		&Fusion&Calibration&Hybrid encoder& ET     & TC     & WT  &Mean   & ET     & TC     & WT  &Mean\\
		\hline
		(1) & & & &0.8351 &0.8753 &0.9080 &0.8728 &15.53 &8.42 &8.37 &10.78\\
		(2) &$\checkmark$ && &0.8720  &0.8848  &0.9278  &0.8949  &9.20  &6.85  &7.18  &7.74\\
		(3) &&$\checkmark$&  &0.8566  &0.8807  &0.9229  &0.8868  &13.96  &6.70  &7.80  &9.49\\
		(4) &&&$\checkmark$ &0.8548  &0.8889  &0.9213  &0.8885  &13.18  &7.41  &7.48  &9.36\\
		(5) &$\checkmark$&&$\checkmark$&0.8767  &0.8827  &0.9283   &0.8959   &7.60  &6.96  &8.24  &7.60\\
		(6) &$\checkmark$&$\checkmark$& &0.8568  &0.8841   &0.9262  &0.8890  &13.39  &9.02  &8.09   &10.17\\
		(7) &&$\checkmark$&$\checkmark$        &0.8783   &0.8989  &0.9314  &0.9029  &10.71  &\textbf{6.36}   &6.70  &7.92\\
		\hline
		(8) &$\checkmark$&$\checkmark$&$\checkmark$  &\textbf{0.8850}  &\textbf{0.9016}  &\textbf{0.9333} &\textbf{0.9066} &\textbf{5.93}  &6.54  &\textbf{6.20} &\textbf{6.22}\\
		\hline
	\end{tabular}
\end{table*}

\subsubsection{Effectiveness of Multi-modal Fusion}
In this paper, we introduce the clinical knowledge-driven formulation by first re-grouping the modalities before feeding them into the model. And then the proposed MCCA block is used to extract the cross-modal features. To prove the effectiveness of the multi-modal fusion strategy, we remove the cross-modal attention in the MCCA blocks.

Compared with Model (1), Model (2) with the clinical knowledge-driven multi-modal fusion strategy achieves an obvious improvement in both Dice score and HD95 distance. It even outperforms most of the SOTA methods shown in Table~\ref{tab:quantitative}. The mean HD95 distance of Model (2) is in the 2nd place (Ours: 6.22, Model (2): 7.74, VTNet: 8.05), while the mean Dice score is in the 3rd place (Ours: 0.9066, Swin UNETR: 0.8984, Model (2): 0.8949). It proves that the multi-modal fusion strategy effectively improves the segmentation results with more precise boundaries. When combining our multi-modal fusion strategy with the hybrid encoder and feature calibration, the segmentation results are further improved.


\subsubsection{Effectiveness of Hybrid Encoder}
In order to introduce both long-range information and inductive bias, we design a hybrid encoder of transformer and CNN. To demonstrate the effectiveness of the hybrid encoder, we compare our model with the one that replaces the hybrid branches in the MCCA block with the swim transformer blocks. 

Quantitative results show that the hybrid encoder (Model (4)) generally improves the segmentation results with better Dice scores compared with the baseline model (Model (1)). However, due to the lack of the multi-modal fusion strategy, HD95 distance of Model (4) is larger than that of Model (2). When combining the multi-modal fusion and hybrid encoder in Model (5), both Dice score (Model (4): 0.8885, Model (5): 0.8959) and HD95 distance (Model (4): 9.36, Model (5): 7.60) are improved. This observation means that the hybrid encoder has a positive impact on the segmentation results while the lack of the multi-modal fusion strategy may lead to imprecise segmentation boundaries.


\subsubsection{Effectiveness of Feature Calibration}
The TCFC block is designed to bridge the gap between the skip connection (transformer) features and the mainstream (CNN) features. Even though the encoder is a hybrid one, the gap between the features of the encoder and the decoder still exists. In this experiment, we disable the feature calibration by replacing the TCFC block with a conventional CNN decoder and directly concatenate the mainstream features and the skip connection features.  

With the hybrid encoder only in Model (3), TCFC can also improve the segmentation results compared with Model (1) without it. Actually, combining transformer with CNN in the encoding phase can introduce the inductive bias, which can also bridge the gap between the features extracted from transformer and CNN. When associating hybrid encoder with feature calibration (TCFC) in Model (7), the segmentation performance is further improved in both Dice score (Model (3): 0.8868, Model (4): 0.8885, Model (7): 0.9029) and HD95 distance (Model (3): 9.49, Model (4): 9.36, Model (7): 7.92).

To summarize the ablation studies, the effectiveness of all the technical novelties has been evaluated. Each one of them alone can improve the brain tumor segmentation performance. Multi-modal fusion mainly serves for finding the correlation between the correlated modalities, resulting in more precise boundaries with the lower HD95 distance. Hybrid encoder leverage the strengths of both transformer and CNN. Associating the hybrid encoder with feature calibration can bridge the gap between transformer and CNN, resulting in more precise segmentation results. The final model with all the technical novelties achieves the best segmentation results quantitatively and qualitatively. 

\begin{figure*}[t]
	\centering
	\includegraphics[width=0.99\linewidth]{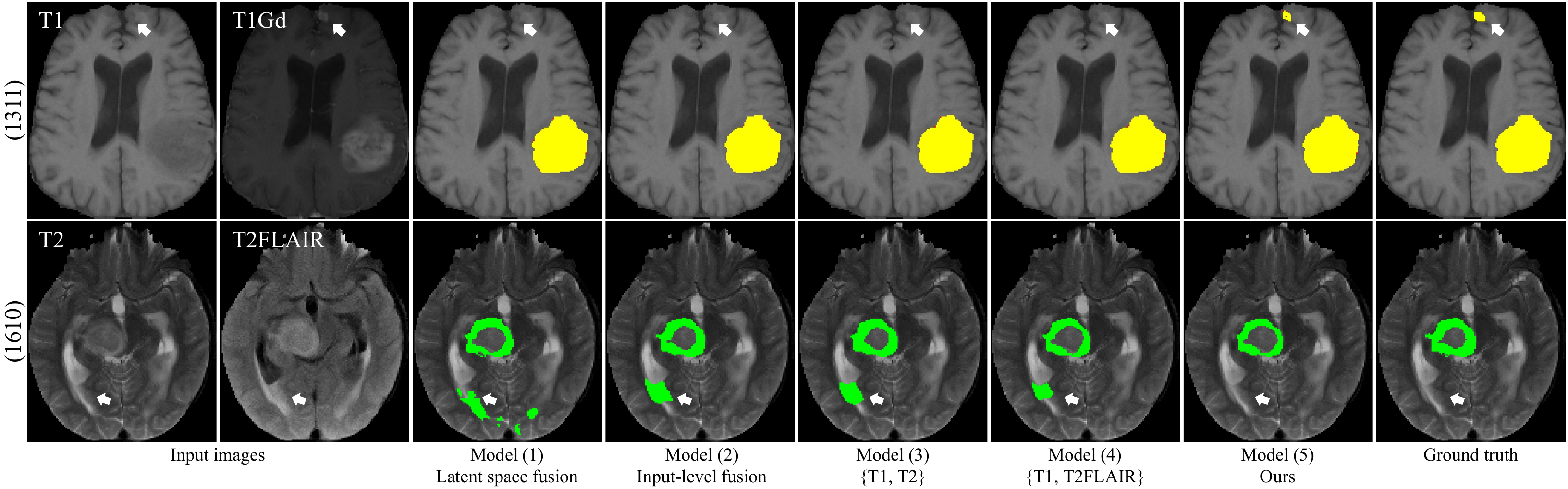}
	\caption{Visualization of the segmentation results in different categories. The top row shows the segmentation results of the model in the enhance tumor and necrotic regions. The bottom row shows the segmentation results of the model in the edema region. We only show one specific categories for better visualization. The leftmost numbers indicate the IDs of the samples.}
	\label{fig:ablation-fusion}
\end{figure*}

\begin{table*}[t]
	\centering
	\caption{Comparison with different multi-modal fusion strategies. Modalities in the same group are marked by the same symbol $\star$ or $\circ$.}
	\label{tab:fusion}
	\begin{tabular}{l|cccc|cccc|cccc}
		\hline 
		\multirow{2}{*}{Models} &\multirow{2}{*}{T1}&\multirow{2}{*}{T1Gd}&\multirow{2}{*}{T2}&\multirow{2}{*}{T2FLAIR}&\multicolumn{4}{c|}{Dice $\uparrow$} & \multicolumn{4}{c}{HD95 $\downarrow$}\\
		&&&&& ET    & TC    & WT  &Mean   & ET     & TC     & WT  &Mean\\
		\hline
		(1) Latent          &\multicolumn{4}{c|}{Feature-level fusion}&0.8783   &0.8989  &0.9314  &0.9029  &10.71  &6.36   &6.70  &7.92\\
		(2) Input         &\multicolumn{4}{c|}{Input-level fusion}&0.8681   &0.8954  &0.9321  &0.8985  &12.39   &\textbf{6.35}   &6.35 &8.37\\ 
		(3) Ours      &$\circ$&$\star$&$\circ$&$\star$&0.8753   &0.8933  &0.9285  &0.8990  &9.30   &6.71   &6.87  &7.63\\
		(4) Ours      &$\circ$&$\star$&$\star$&$\circ$&0.8786   &0.8865  &0.9292  &0.8981  &7.40   &6.93   &7.03  &7.12\\
		(5) Ours*   &$\circ$&$\circ$&$\star$&$\star$&\textbf{0.8850}  &\textbf{0.9016}  &\textbf{0.9333} &\textbf{0.9066} &\textbf{5.93}  &6.54  &\textbf{6.20} &\textbf{6.22}\\
		\hline
	\end{tabular}
\end{table*}

\subsection{Comparison with Different Fusion Strategies}
Furthermore, we also compare our model with different multi-modal fusion strategies in Table~\ref{tab:fusion}. (1) In this model, we remove the cross-modal attention in the MCCA block. Each modality has an individual branch to extract the features. The latent features are concatenated before feeding into the bottleneck layer. (2) We directly concatenate four modalities at the input level and with only one single transformer branch in the encoder. (3)-(4) The model architecture remains unchanged with different groups of the input modalities. The modalities in the same group are marked by the symbols $\star$ and $\circ$. (5) Ours marked by * is the final model.

As shown in Table~\ref{tab:fusion}, two conventional ways of feature-level fusion and input-level fusion can both achieve reasonable results in Dice score. But they are less effective in HD95 distance. Since HD95 distance focuses more on the correctness of the contours. Model (3) and Model (4) with different groups of modalities can also improve the precision of the contours with lower HD95 distance. It reveals that the proposed MCCA block with cross-modal attention is superior to conventional feature-level fusion and input-level fusion ways. MCCA block is easier to find the correlation between two modalities, even they are less correlated. Inspired by the clinical knowledge from the radiologists, grouping two highly correlated modalities (T1 and T1Gd) in Model (5) achieves the best performance, especially for the enhanced tumor (ET) with DICE of 0.8850 and HD95 of 5.93. Because radiologists define the enhanced tumor by comparing the T1 and T1Gd modalities.

Fig.~\ref{fig:ablation-fusion} demonstrates the qualitative results of this experiment. We show two distinct cases to visualize the segmentation results which get benefit from the clinical knowledge-driven design. In this figure, we only show the most relevant category in the results for clearer visualization. In the first row, we show the enhanced tumor (ET) results and the corresponding T1 and T1Gd modalities. As we mentioned above, radiologists assess ET by comparing the T1 and T1Gd. Following this clinical knowledge, our multi-modal fusion strategy successfully help the model find out another small lesion pointed by the arrow. While the models with the other fusion strategies or inappropriate input combinations miss this lesion. The second row show the case of edema, which is assessed by comparing the T2 and T2FLAIR modalities. Our model with a reasonable multi-modal fusion way can suppress the false positive regions pointed by the arrow. 

\section{Conclusion}
In this paper, we propose a novel clinical knowledge-driven brain tumor segmentation model with four input MRI modalities, named CKD-TransBTS. We deeply analyze the way how radiologists assess and diagnose brain tumor from multiple modalities and introduce the clinical knowledge into our multi-modal fusion strategy. By simply re-grouping the input modalities according to the imaging principles of them, our model can get obvious improvement as shown in the quantitative and qualitative experiments. This multi-modal fusion strategy effectively improves the segmentation results with more precise boundaries and suppresses the false positive. We believe that introducing some relevant prior knowledge or the clinical knowledge into the model design could be an very effective way to benefit the model representation and let model learn the inherent characteristics of the data beyond the labels.

In the technical perspective, we leverage the strengths of both transformer and CNN by proposing a hybrid transformer model. In the encoder phase, we propose a novel Modality-Correlated Cross-Attention (MCCA) block to fuse and extract the multi-modal features. In the decoding phase, a Trans\&CNN Feature Calibration (TCFC) block is proposed to bridge the gap and the bias between the features of transformer and CNN. The effectiveness of all the technical novelties have been evaluated by the experiments.

%
\bibliography{ref}
\bibliographystyle{IEEEtran}

\end{document}